\documentclass[conf, onecolumn]{ao4elt3}
\usepackage{graphics}  
\usepackage{amsmath}
\usepackage[varg]{txfonts} 
\usepackage[latin1]{inputenc}

\begin{document}

\def\refjnl#1{{\rmfamily#1}}%
\newcommand\mnras{\refjnl{MNRAS}}%
\newcommand\apj{\refjnl{ApJ}}%
\newcommand\pasp{\refjnl{PASP}}%

\pagestyle{plain}
\title{SUPER-RESOLUTION IMAGING WITH AN ELT: KERNEL-PHASE INTERFEROMETRY}
\author{Frantz Martinache\inst{1}\thanks{frantz.martinache@oca.eu}}
\institute{Observatoire de la C\^ote d'Azur, Bd de l'Observatoire,
  06304 Nice, France}
\abstract{
Kernel-phase is a recently developed paradigm that tackles the
classical problem of image deconvolution, based on an interferometric
point of view of image formation. Kernel-phase inherits and borrows
from the notion of closure-phase, especially as it is used in the
context of non-redundant Fizeau interferometry, but extends its
application to pupils of arbitrary shape, for diffraction limited
images. 
The additional calibration brought by kernel-phase boosts the
resolution of conventional images and enables the detection of
otherwise hidden faint features at the resolution limit and beyond, a
regime often refered to as super-resolution, which for a 30-meter
telescope in the near IR, this translates into a resolving power
smaller than 10 mas. Kernel-phase analysis of archival space and
ground based AO data leads to new discoveries and/or improved relative
astrometry and photometry. The paper presents the current status of
the technique and some of its recent developments and applications
that lead to recommendations for super-resolution imaging with ELTs.
}
\maketitle
\section{Introduction}
\label{intro}

Imaging is the starting point of a lot of astronomical investigation:
the image is the place where observers identify new sources, track
their position relative to other reference points 
and/or follow their evolution as a function of time and/or
wavelength.
The image is indeed a remarkable locus in optics, where the photons of
multiple sources are optimally segregated. Within the isoplanetic
field of the instrument used for the acquisition, this image can be
described as the result of the following convolution product:

\begin{equation}
  I = O \otimes \mathrm{PSF},
  \label{eq:convol}
\end{equation}

\noindent
where $I$, the distribution of intensity measured in the image, is a
representation of the true structure of the object $O$, modified by
the optical system point spread function $\mathrm{PSF}$.
Imaging in the high-angular resolution regime essentially comes down
to attempting to solve this deconvolution problem, and separate the
two terms $I$ and $\mathrm{PSF}$. The problem is unfortunately
ill-posed, with too many unknown for too few constraints.
Adaptive optics (AO) obviously plays an important part in optimizing
this segregation of photons, by concentrating the light of each source
otherwise scattered over a large halo, into a narrow
diffraction-limited pattern, that facilitates their identification.

Yet even at the highest Strehl, this apparently optimal segregation of
photons is often not sufficient in solving some important problems
such as: (1) the identification of faint sources or structures in the
direct neighborhood of a bright object: in this context, the faint
source one tries to detect is competing for the observer's attention
with the diffraction features of its host or (2) the discrimination of
sources of comparable brightness so close to each other that they are
said non-resolved.

Even if they are small, residual time-variable aberrations are
responsible for the presence of a combination of static, quasi-static
and rapidly evolving speckles in the image. 
Different schemes have been developed to sort out the PSF from the
object function, and a fairly successful approach is to use some form
of diversity in the PSF: angular differential imaging
\cite{2006ApJ...641..556M} for instance, uses field-rotation to
differentiate true companions from quasi-static speckles. If enough
field-rotation occurs over a time that is less than the characteristic
quasi-static aberration time-scale, the diversity between the
orientation of the sky and the quasi-static PSF allows the calibration
of the PSF, leading to greatly enhanced detection limits. The
calibration requires that the angular rotation induces sufficient
local linear displacement of the genuine object features in order to
avoid self-subtraction, which makes the technique little relevant for
angular separations smaller than 0.5''.
At small angular resolution, active alternatives are becoming
available, with updated AO systems employing additional active optics
to introduce wavefront diversity, to create higher-contrast regions in
the image using techniques such as speckle nulling
\cite{2012PASP..124.1288M}. 

This paper presents an alternative approach, that looks at image
formation from an interferometric point of view. Instead of focusing
on the image, one examines its Fourier-transform counterpart. The
approach builds upon the recent adaptation of non-redundant aperture
masking interferometry to AO corrected images
\cite{2006SPIE.6272E.103T}. Non-redundant masking interferometry takes
advantage of the self-calibration properties of an observable quantity
called the closure-phase \cite{1958MNRAS.118..276J}. Closure-phase
extracted from images acquired with a sparse aperture mask are indeed
immune to residual wavefront of low spatial frequency, characterized
by scale larger than the size of the sub-apertures, making them a
powerful observable for high contrast imaging
\cite{2012ApJ...745....5K}.

It was demonstrated that under high-Strehl conditions, the notion of
closure-phase, requiring non-redundant aperture, can be generalized to
arbitrarily shaped (i.e. including redundant) pupils
\cite{2010ApJ...724..464M}. Instead of trying to solve the ill-posed
problem of image deconvolution posed in Eq. \ref{eq:convol}, the
approach presented here allows to extract from conventional images
observable quantities called kernel-phases that just like
closure-phase, are immune to residual wavefront aberrations, including
non-common path errors.

Section \ref{sec:examples} introduces some simple examples that
describe how one can generalize the original idea of closure-phase
to make it compatible with redundant arrays. Section \ref{sec:kpao}
explains how new observables called kernel-phases can be extracted
from conventional well corrected images and Section \ref{sec:appli}
presents some applications.

\section{Beyond the closure-phase in simple cases}
\label{sec:examples}

\begin{figure}[!ht]
\centerline{\resizebox{0.8\columnwidth}{!}{\includegraphics{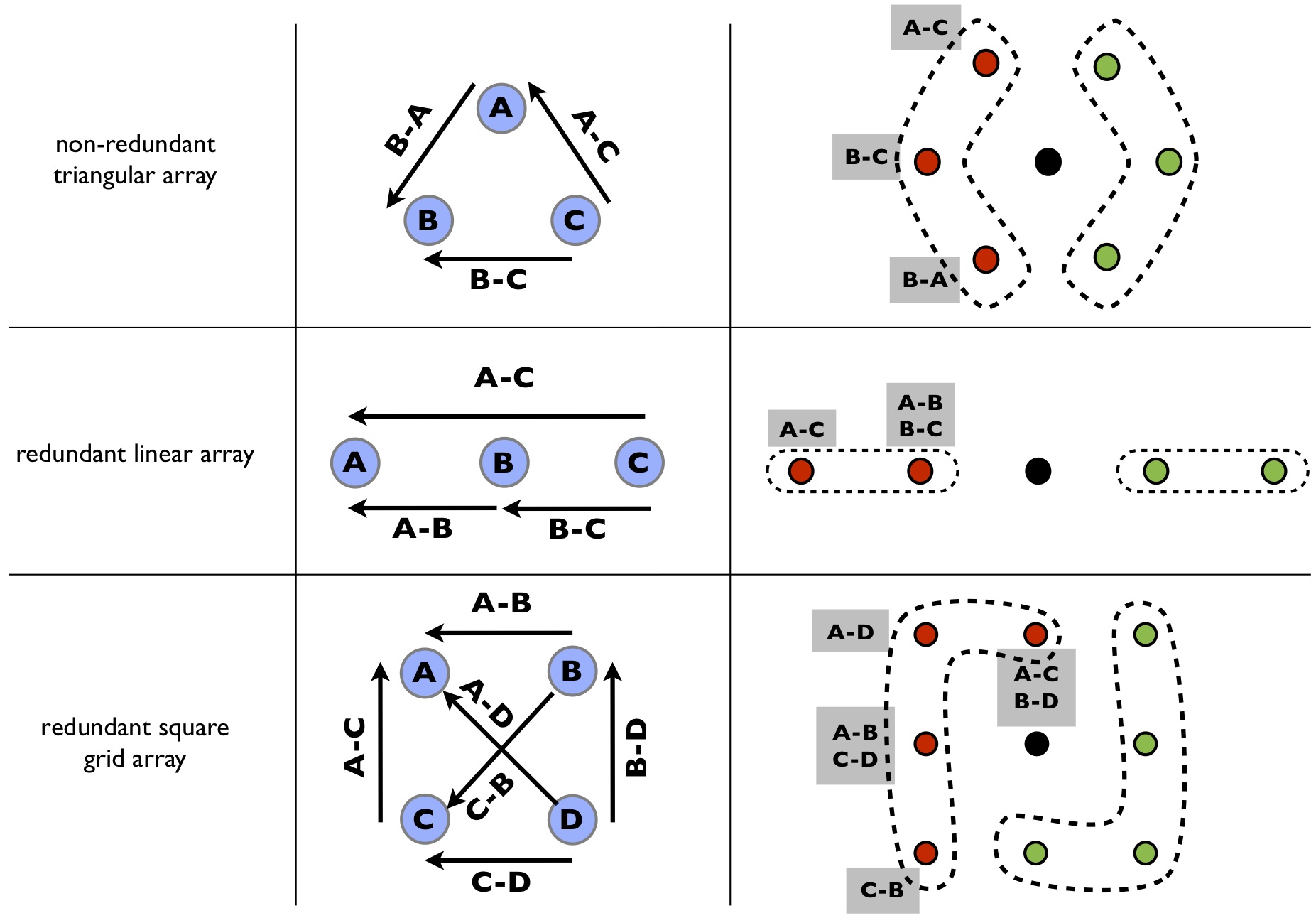} }}
\caption{
  Pupil and corresponding $(u,v)$-coverage for three of the simplest
  possible sparse pupil geometries. From top to bottom: (1)
  non-redundant triangular array; (2) redundant linear array and (3)
  redundant square grid.
}
\label{f:geom}
\end{figure}

Although the final application of this work is relevant to classical
imaging with a conventional (i.e. non-sparse) telescope aperture, it
is useful to go look at several sparse geometries to understand the
basis for the model used for kernel-phase.
In that scope, Fig. \ref{f:geom} features three of the simplest
possible sparse pupil geometries, for which we examine the way the
wavefront aberrations propage into the Fourier-transform of images.
Although apertures are represented here with a non-negligible diameter
for readability, in practice, they are considered to be
point-like. For all geometries, Fig. \ref{f:geom} labels apertures
with letters and baselines with couple of letters.

The first geometry, is a non-redundant 3-aperture array.
Apertures are labeled $\mathrm{A}$, $\mathrm{B}$ and $\mathrm{C}$, and
form three distinct baselines $\mathrm{A-C}$, $\mathrm{B-C}$ and
$\mathrm{B-A}$. An image being a real function, its Fourier-transform
exhibits some standard parity properties: it is even in amplitude and
odd in phase, explaining why each baseline translates into a pair of
points in the $(u,v)$-plane: only one of point per baseline needs to
be considered: they are highlighted in red while their symmetric is in
green.
This work is for now restricted to the phase, and will ignore
amplitude effects entirely. Each baseline gives access to a distinct
measurement of the phase of the target of interest, a 3-component
vector noted $\Phi_O$.
This phase is however polluted by the phase delay along the baseline
formed by the two apertures. The three equations for the different
baselines phase simply write as:

\begin{eqnarray*}
  \Phi^{BC} &=& \Phi_O^{BC} + (\varphi_B-\varphi_C) \\
  \Phi^{AC} &=& \Phi_O^{AC} + (\varphi_A-\varphi_C) \\
  \Phi^{BA} &=& \Phi_O^{BA} + (\varphi_B-\varphi_A).
\end{eqnarray*}

It doesn't take much to group these three equations into a compact
matrix form:

\begin{eqnarray*}
  \Phi &=& \Phi_O + \mathbf{A} \cdot \varphi, \\
  \mathbf{A} &=& \begin{pmatrix} 0 & 1 & -1 \\ 1 & 0 & -1 \\ -1 & 1 & 0 \end{pmatrix} 
  \label{eq:triangl}
\end{eqnarray*}

\noindent
with $\mathbf{A}$, referred to as the phase transfer matrix, encoding
the information about the array. This geometry is often used to
introduce the idea of closure-phase \cite{2000plbs.conf..203M} in
long-baseline interferometry. The sum of the phases measured along
baselines forming a closing triangle: $\mathrm{C-B}$, $\mathrm{B-A}$,
$\mathrm{A-C}$, is immune to the phase delay term $\varphi$. One can
write this closure-phase relation in terms of a left-hand operator
$\mathbf{K} = \begin{pmatrix} -1 & 1 & 1\end{pmatrix}$, that verifies
  $\mathbf{K} \cdot \mathbf{A} = \mathbf{0}$.

The second scenario presented in Fig. \ref{f:geom} is not as
straightforrward, since one of the apertures of the previous array was
moved so as to form a redundant baseline: $\mathrm{A-B}$ and
$\mathrm{B-C}$ are indeed identical, while the baseline $\mathrm{A-C}$
remains unique.
In the $(u,v)$-plane, the contributions of redundant baselines add up
in the same place, resulting in two Fourier-phase components instead
of three.
Two baselines measure the same target phase object component $\Phi_O$,
however each shifted by an amount that depends on the baseline phase
delay. The resulting phase in this point is the argument of the sum of
two phasors:

\begin{equation}
\Phi = \mathrm{Arg}[\exp{i(\Phi_O+\varphi_A-\varphi_B)} +
\exp{i(\Phi_O+\varphi_B-\varphi_C)}].
\label{eq:arg}
\end{equation}

For large phase delays (i.e. wavefront aberrations), this resulting
phase can take any value between 0 and 2$\pi$. However if one assumes
that phase errors are small ($\Delta\varphi \lesssim 1$ 
radian), which corresponds to a reasonable requirement on the AO
system, Eq. \ref{eq:arg} can be linearized and considerably
simplifies as:

\begin{equation*}
  \Phi = \Phi_O + \frac{1}{2} (\varphi_A-\varphi_C).
\end{equation*}

Just like for the triangular geometry, the linear equations for the
phase can be gathered into a compact matrix form:

\begin{equation}
\Phi = \Phi_O + \mathbf{R}^{-1} \cdot \mathbf{A} \cdot \varphi,
\label{eq:generic}
\end{equation}

\noindent
with:
\begin{equation*}
\mathbf{R}^{-1} = \begin{pmatrix} 1 & 0 \\ 1 &
  \frac{1}{2} \end{pmatrix} \mathrm{~~and~~} \mathbf{A}
=\begin{pmatrix} 1 & 0 & -1 \\ 1 & 0 & -1 \end{pmatrix}.
\end{equation*}

For this array, the only possible closure relation is $\mathbf{K}
= \begin{pmatrix} 1 & -2 \end{pmatrix}$. Here, the phase transfer
matrix has been split into two components $\mathbf{R}^{-1}$ and
$\mathbf{A}$, that respectively encode the redundancy of the baselines
and the phase relationships.

With the last two examples in mind, and assuming that the
linearization of Eq. \ref{eq:arg} holds for arbitrarily redundant
arrays, the last example is not difficult to write down directly using
the same Eq. \ref{eq:generic}, with the following matrices:

\begin{equation*}
\mathbf{R}^{-1} = 
\begin{pmatrix} 
  1 &      0     & 0 & 0 \\ 
  0 &\frac{1}{2} & 0 & 0 \\
  0 &      0     & 1 & 0 \\
  0 &      0     & 0 & \frac{1}{2}
\end{pmatrix} 
\mathrm{~~~and~~~}
\mathbf{A} = 
\begin{pmatrix} 
  1   &   0   &   0   &   -1 \\ 
  1   &  -1   &   1   &   -1 \\
  0   &  -1   &   1   &    0 \\
  1   &   1   &  -1   &   -1
\end{pmatrix}. 
\end{equation*}

Two simple closure relations can be found by hand, and can again be
summarized by a left-hand operator $\mathbf{K}$, so that:

\begin{equation*}
  \mathbf{K} = 
  \begin{pmatrix} 
    1   &   -2   &   1   &   0 \\ 
    1   &  0   &   -1   &   -2
  \end{pmatrix}. 
\end{equation*}

\section{Kernel-phase}
\label{sec:kpao}

The three examples introduced in Section \ref{sec:examples} suggest
that, in the low-aberration (high-Strehl) regime, it is possible to
identify, in the Fourier-phase space, a linear equivalent to the
image-object convolution relation of eq. \ref{eq:convol}, of the form
shown in eq. \ref{eq:generic}, even when the aperture is redundant.
The most important element of this model is the phase transfer matrix
$\mathbf{A}$, describing the way the pupil wavefront aberrations
propagate into spurious phase in the Fourier-plane. The diagonal
matrix $\mathbf{R}^{-1}$ is a scaling operator, that contains the
inverse value of the baseline redundancy.
$\mathbf{A}$ and $\mathbf{R}^{-1}$ can of course be combined into a
single operator.

\begin{figure}
  \centerline{\resizebox{0.8\columnwidth}{!}{\includegraphics{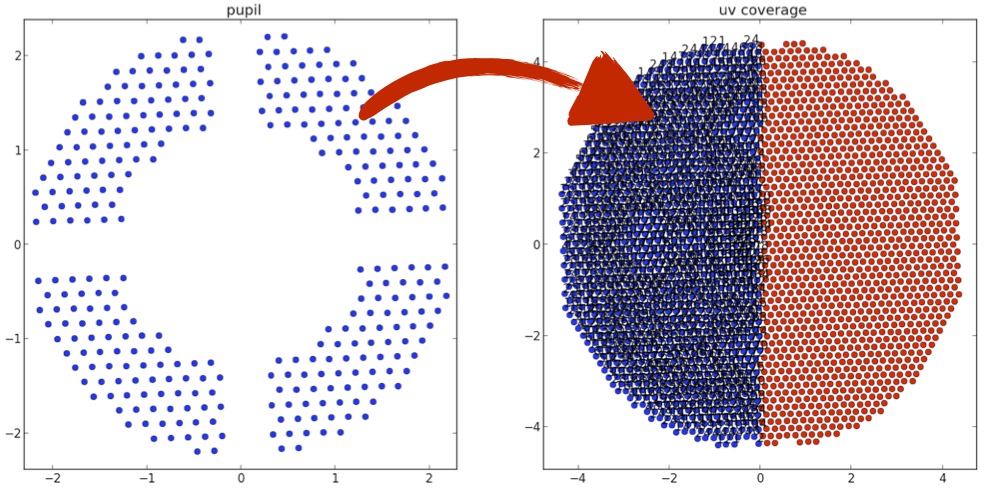} }}

  \caption{
    Example of discrete model that allows to construct the operator
    $\mathbf{A}$ that describes the way the pupil phase $\varphi$
    propagates into the Fourier-plane (cf. Eq. \ref{eq:generic}). 
    Left: discrete model of the pupil of the instrument, following a
    regular hexagonal grid. 
    Right: the resulting distribution of spatial frequencies sampled by
    this geometry.
  }
  \label{f:palomar}
\end{figure}

To work on data acquired with a conventional telescope aperture, one
must start by building a discrete model of the pupil.
Figure \ref{f:palomar} shows one working example for the ``medium
cross pupil'' of the Palomar Hale Telescope PHARO instrument
\cite{2001PASP..113..105H}. The discrete model of the instrument
pupil, uses a regular grid pattern, whose density is representative of
the continuous pupil.

The model in Fig. \ref{f:palomar} projects 332 pupil elements onto a
grid of 1128 distinct sample points in the Fourier domain. The system
is obviously large enough to justify delegating the identification of
the kernel-phase operator $\mathbf{K}$ to an automated algorithm. One
possible solution can be found as one of the products of the singular
value decomposition (SVD) or the phase transfer matrix
$\mathbf{A}$. The total number of kernel-phase relations $n_K$ is
given by the number of zero singular values of $\mathbf{A}$. If the
SVD of $\mathbf{A}$ writes as: $\mathbf{A} = \mathbf{U} \mathbf{W}
\mathbf{V}^T$, where $\mathbf{W}$ is a diagonal matrix containing
the singular values of $\mathbf{A}$, and $\mathbf{U}$ and
$\mathbf{V}^T$ are unitary matrices, one possible set of orthogonal
kernel-phase relations can be found in the columns of $\mathbf{U}$
that correspond to zeros on the diagonal of $\mathbf{W}$. They form an
orthonormal basis for the left null space (or kernel) of the phase
transfer matrix, hence the name kernel-phase.

Gathered into the operator $\mathbf{K}$, these relations are then
applied to the phase measured in the Fourier plane $\Phi$, to extract
information about the target of interest that is immune to residual
wavefront error $\varphi$:

\begin{equation}
\mathbf{K} \cdot \Phi = \mathbf{K} \cdot \Phi_O.
\end{equation}

Just like with closure phase, that collapses several phase
measurements into a reduced number of observable quantities, some of
the phase information seems lost in the process: these high quality
can however be used as constraints for an interferometric imaging
algorithm, or a parametric model.
The SVD of the phase transfer matrix for the model shown in
Fig. \ref{f:palomar} shows that 962 kernel-phases can be extracted
from any single image, which means that 85~\% of the phase information
is directly recoverable.

Once the paving of the $(u,v)$-plane and the matching kernel-phase
relations are identified, they are saved in a template and used for
extracting the phase information from the data. Before being
Fourier-transformed, frames undergo traditional dark subtraction
and flat-fielding procedure. 
To limit the impact of detector readout noise, the data can can be
windowed, for instance with a ``super-Gaussian'' ($\exp{-(r/r_0)^4}$)
radial profile.
After the frame is Fourier-transformed, the phase is sampled at the
relevant $(u,v)$ coordinates and assembled into the vector
$\Phi$. Assuming that the data is at least Nyquist-sampled allows all
spatial frequencies to be extracted.
Kernel-phase observables $\mathbf{K}\Phi$ are constructed using the
pre-determined relations for each frame.
Multiple frames on a given target and/or the availability of frames
acquired on single stars allow further characterization of the
Ker-phase data, using statistics and/or additional calibration.

\section{Applications}
\label{sec:appli}

\subsection{HST/NICMOS}

One of the most productive use of closure-phase with non-redundant
masking interferometry with AO has so far been the search for faint
companions around a wide variety of targets 
\cite{2008ApJ...679..762K,2011ApJ...731....8K,2008ApJ...678..463I,2012ApJ...745....5K}. Since kernel-phase takes advantage of the same
self-calibrating properties as closure-phase, one expects kernel-phase
analysis of high-Strehl ratio images to cover a similar fraction of
the parameter space, extending from $\sim0.5\lambda/D$ to a few
$\lambda/D$.
Not-saturated images acquired in the near-infrared by NICMOS onboard
HST offer the ideal test case for this technique: the Strehl is quite
high, and image quality is excellent, with sampling better than
Nyquist.

\begin{figure}
  \centerline{
    \resizebox{0.49\columnwidth}{!}{\includegraphics{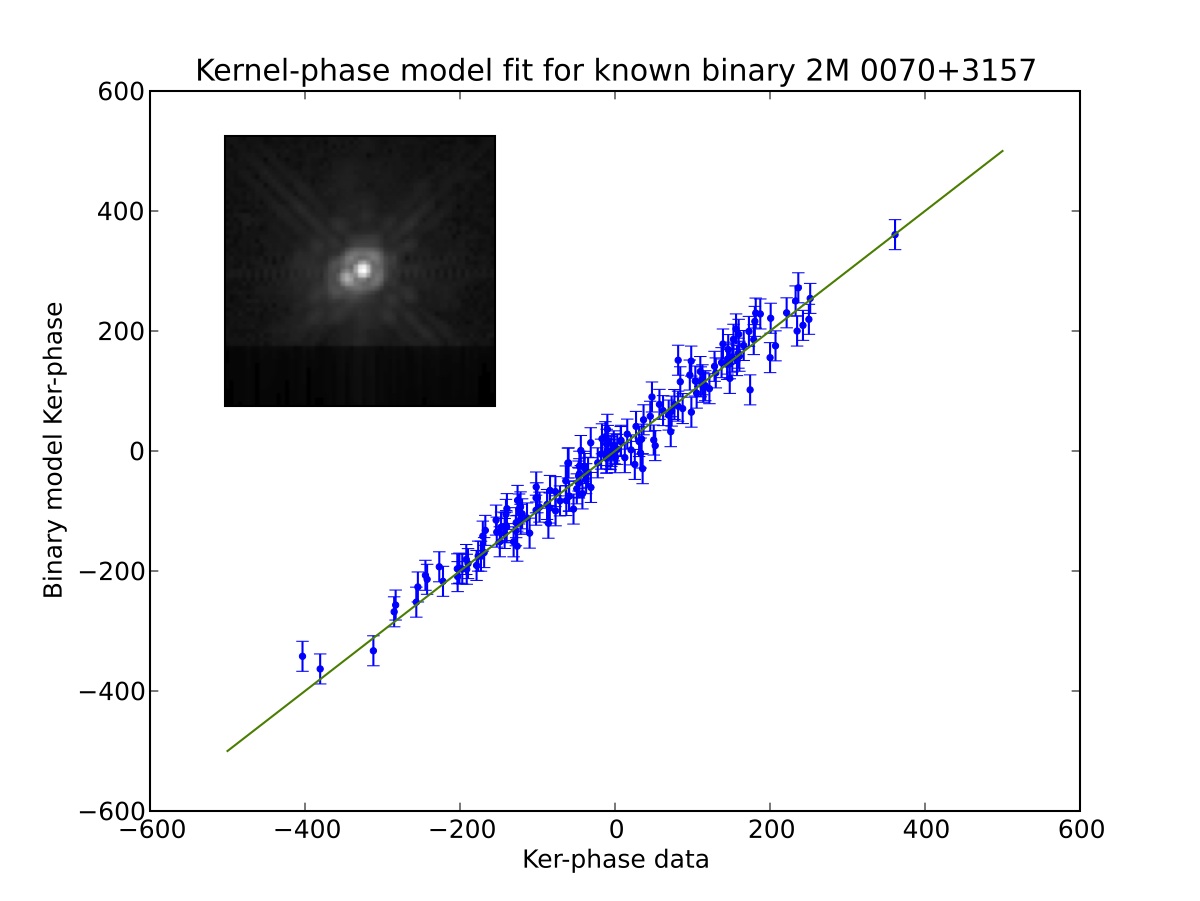}}
    \resizebox{0.49\columnwidth}{!}{\includegraphics{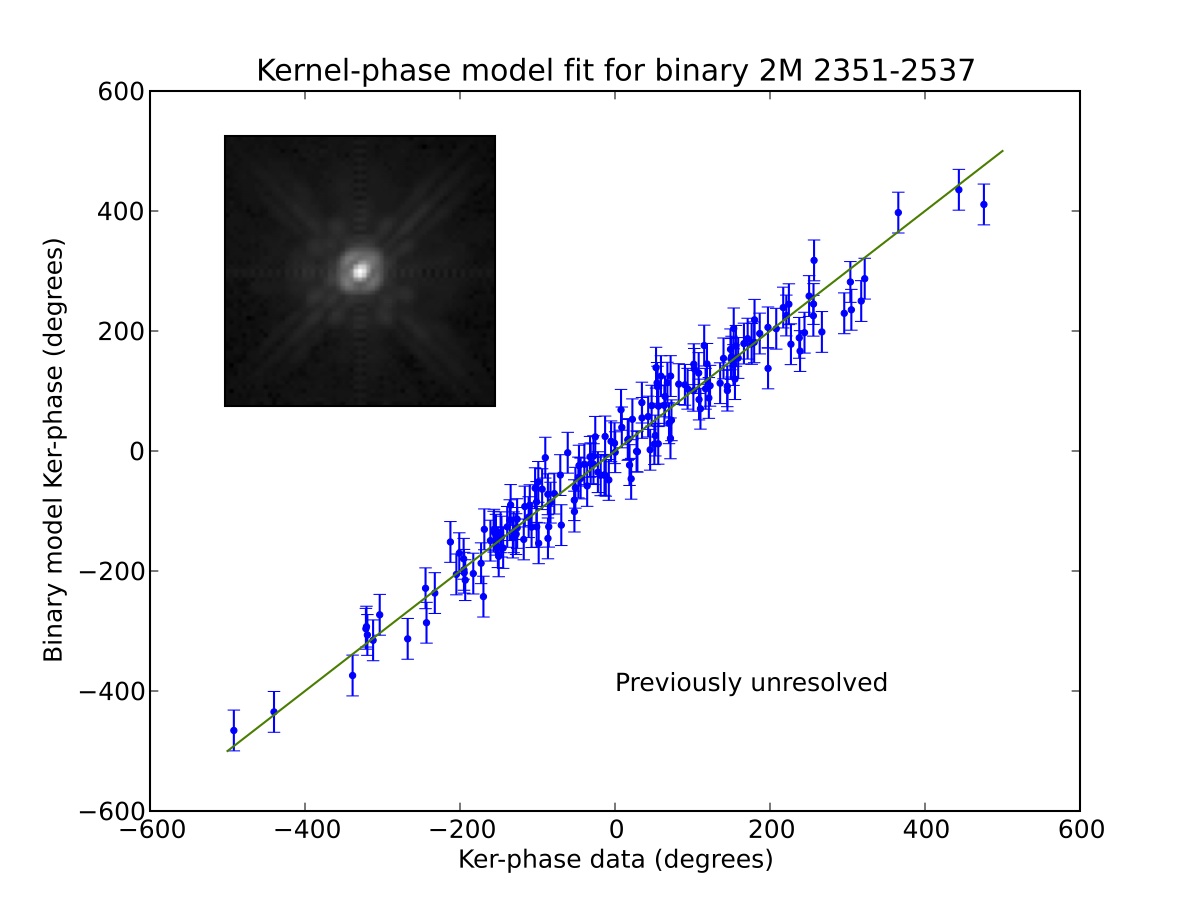}}
  }
  \caption{
    Examples of kernel-phase binary fit results on two ultracool
    dwarfs observed with HST/NIC1 in the F110W filter. 
    Left: 2M 0070+3157 is a known binary, for which the companion was
    discovered from visual examination of the image. Right: 2M
    2351-2537 is a previously unknown binary, that cannot be resolved
    in the direct image. 
    The kernel-phase analysis produces a signal of comparable magnitude
    in either case.
  }
  \label{f:hst}
\end{figure}

A systematic revisit of a homogeneous sample of $\sim$80 nearby
L-dwarfs previously observed with HST/NICMOS was performed by
\cite{2013ApJ...767..110P} in two filters (F110W and F170M) using the
kernel-phase approach outlined in this paper.
This study permitted a search for companions down to projected
separations of $\sim$1 AU, revealed new binary candidates and provided
improved relative astrometry for all known binaries.

After extraction, the kernel-phases are used as constraints in a
3-parameter binary model (separation, position angle and
contrast). Conventional likelihood analysis and/or Monte-Carlo
simulations provide a binary solution or contrast detection limits.
Figure \ref{f:hst} shows two examples of kernel-phase data sets
plotted against their best fit binary model.

\subsection{Ground-based AO}

\begin{figure}
  \centerline{
    \resizebox{0.9\columnwidth}{!}{\includegraphics{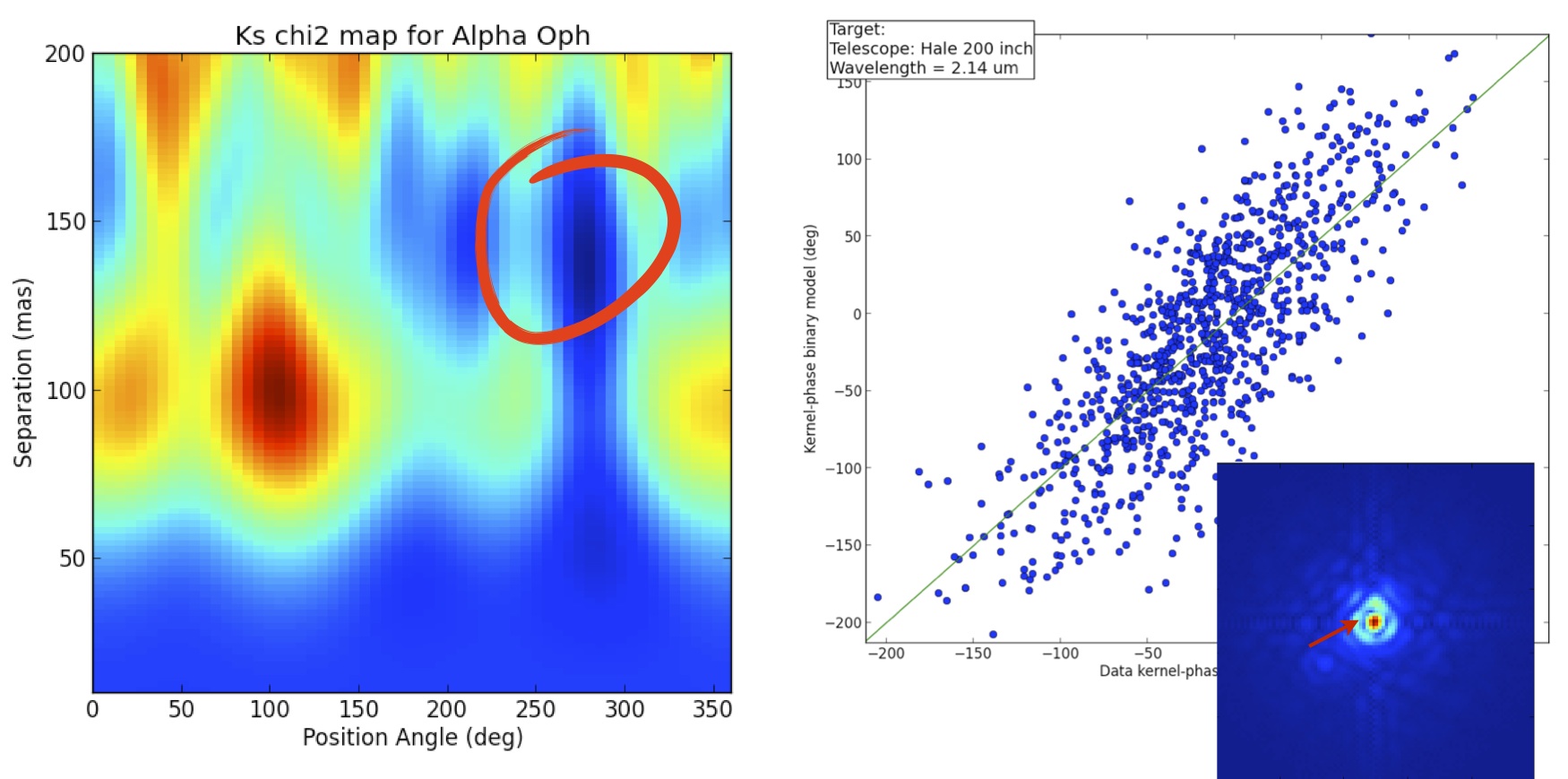}}
  }
  \caption{
    Example of kernel-phase result obtained on ground based AO
    data. Left: map of the $\chi^2$ in the position angle - angular
    separation space for $\alpha$-Ophiucus, observed with PHARO at the
    Palomar Hale Telescope, using the model shown in
    Fig. \ref{f:palomar}.
    A red circle highlights the location of the minimum $\chi^2$.
    Right: correlation plot between the kernel-phase data and the
    binary model for the corresponding location in the $\chi^2$
    space. The image snippet shows (red arrow) that the 30:1 companion
    is apparently invisible, hidden under the first diffraction ring.
  }
  \label{f:alpoph}
\end{figure}

Application of the method is however not restricted to HST/NICMOS
images and can also be applied to ground based AO data, assuming that
the on-axis AO correction produces a Strehl better than $\sim$50~\%.
Fig. \ref{f:alpoph} showcases one example of result achieved using the
model of the Palomar Hale PHARO medium cross pupil introduced in
Sec. \ref{sec:kpao}.

The target, $\alpha$-Ophiucus, is a well known binary with a well
characterized, eccentric orbit with an 8.6 year period
\cite{2011ApJ...726..104H}.
The kernel-phase analysis of multiple PHARO frames acquired in the
K-band revealed the presence of the 30:1 contrast companion at a
position angle 274.6$^{\circ}$, but at an angular separation 136.1
mas ($\sim1.5\lambda/D$), that is directly underneath the first
diffraction ring.
The companion, invisible in the direct image, is clearly detected
using this approach. The position is in very good accordance with the
ephemerides of the orbit.

\section{Interferometric imaging with an ELT}
\label{sec:elt}

The self-calibrating properties of interferometric observables like
closure-phase and kernel-phase make it possible to produce images of
sources, with resolving power better than what the formal diffraction
limit ($\lambda/D$) dictates: a regime called super-resolution.
Well known examples include images of complex sources such as pinwheel
nebulae \cite{2008ApJ...675..698T} obtained by masking the Keck
Telescope with a partially redundant annular mask.
For an ELT with a 30-meter baseline diameter observing in the near IR
($\lambda=1.6 \mu m$), imaging with a resolution of $\sim5$ mas or
less offers fantatic possibilities over a wide range of astrophysical
science cases.

For the most part, the design of interferometric masks for imaging has
been guided by the widely used designs of Golay
\cite{1971JOSA..61..272}. These designs provide a series of compact
solutions using three-fold symmetry that maximize the $(u,v)$-coverage
while ensuring non-redundancy.

$(u,v)$-coverage is indeed an important factor for successful imaging
of complex sources with an interferometer. While it is always possible
to increase this coverage with aperture synthesis techniques, the
discussion is restricted to the analysis of a single snapshot Fizeau
image, such as one acquired with a mask in the pupil of an instrument
behind AO. The imaging capability is entirely dictated by the number
of independent observables (visibilities and kernel-phases) that can
be extracted from the snapshot.

Because it enables extraction of high-quality observables even for
redundant apertures, the notion of kernel-phase allows to go beyond
the rules of Golay, and has the potential to offer solutions for the
imaging of complex sources.
A preliminary comparative study of several designs
\cite{2012SPIE.8445E..04M} has shown that even when they provide
identical $(u,v)$-coverage, some configurations do provide a better
phase information recovery rate (the ratio $n_K/n_{UV}$).

\begin{figure}
  \centerline{
    \resizebox{0.45\columnwidth}{!}{\includegraphics{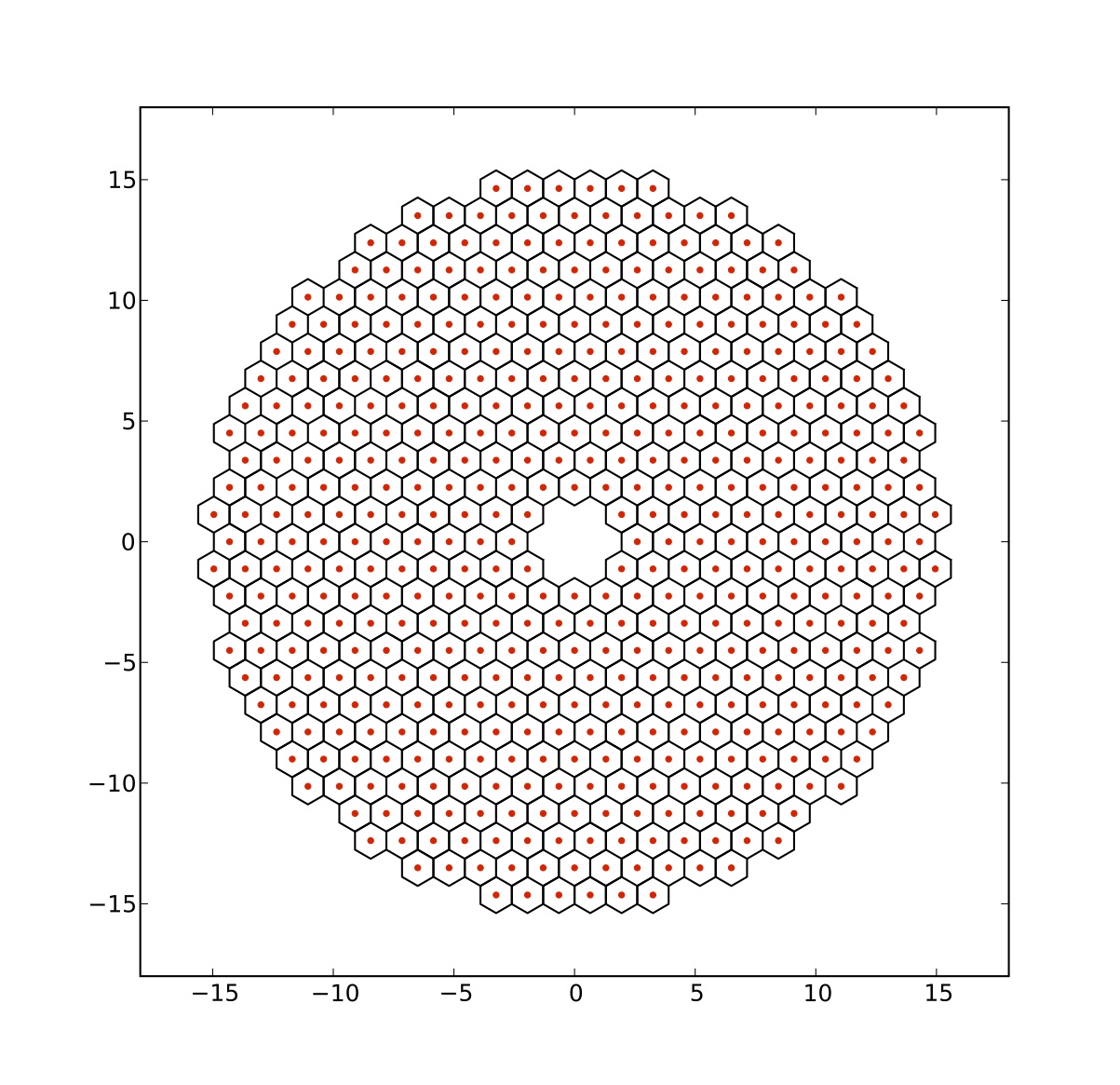}}
    \resizebox{0.45\columnwidth}{!}{\includegraphics{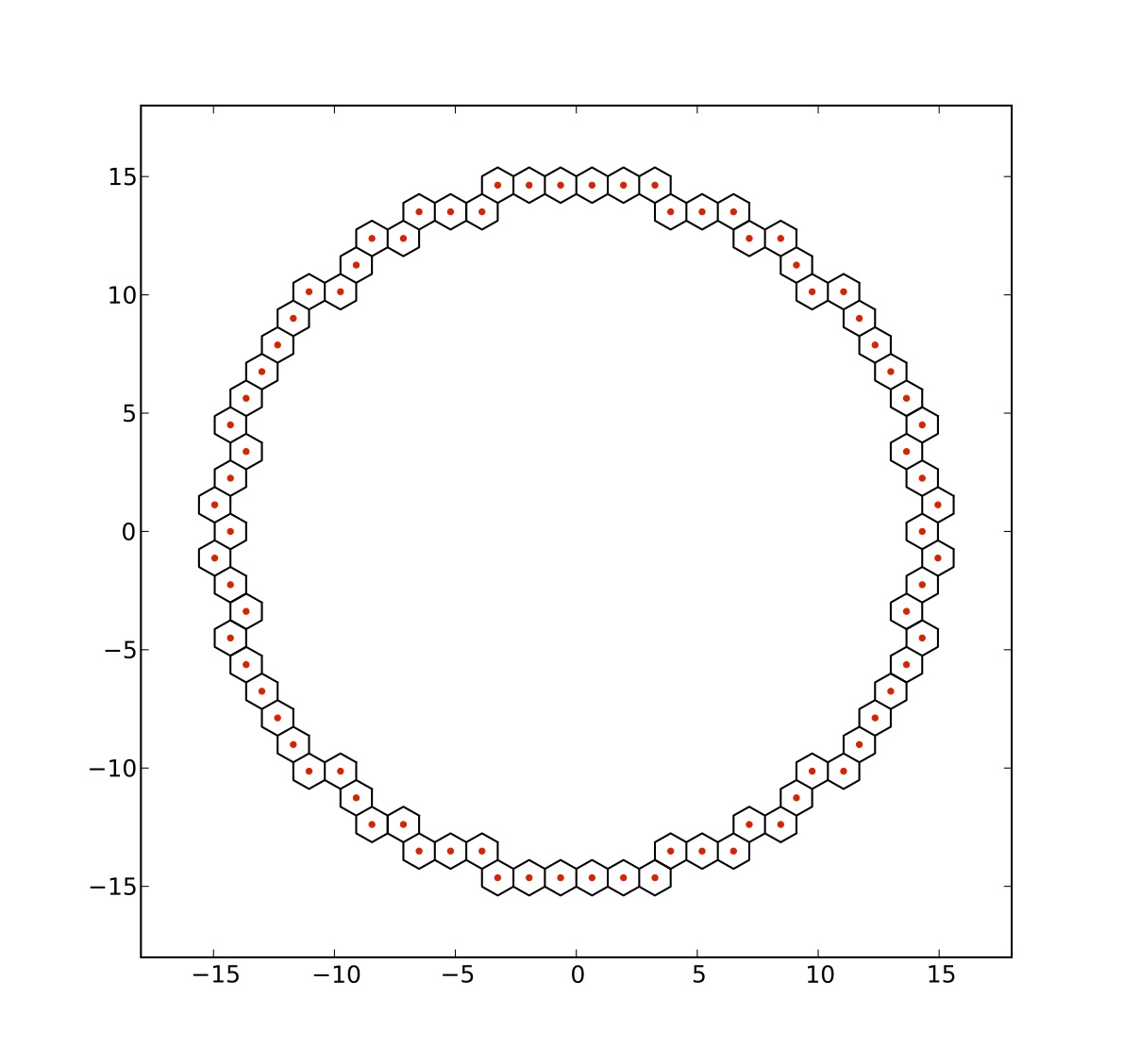}}
  }
  \caption{
    Left: geometry for the 492 segments of the Thirty Meter Telescope
    (TMT).
    Right: geometry of the 78 outermost segments that give access to
    the same $(u,v)$-coverage, but exhibits a better phase information
    recovery rate.
  }
  \label{f:ring}
\end{figure}

What comes out of this study is that an annular configuration does
seem to provide an optimum, with a $(u,v)$-coverage identical to the
full-pupil, but a better yield in terms of phase recovery.
The example shown in Fig. \ref{f:ring} compares, for the Thirty Meter
Telescope (TMT), the full 492-segment aperture to a subset made of the
outermost ring of segments (78).
Going through the analysis highlighted in Section \ref{sec:kpao} for
both configurations reveals that both configurations provide access to
the same $n_{UV} = 972$ distinct sample points in the Fourier
plane. Looking for the number of singular values for the phase
transfer matrix of these apertures shows that for the full aperture,
$n_K=726$ kernel-phase relations are available, giving a reasonable
phase information recovery rate $\sim75~\%$. The annular configuration
exhibits a larger $n_K=933$ number of kernel-phase relations, that is
a total phase information recovery rate $\sim96~\%$.

Of course, the total throughput of the annular configuration is less
favorable ($\sim16~\%$), but remains reasonable in comparison with
most non-redundant masks used with great success for imaging purpose
today on 8-to-10 meter telescopes.
Moreover, because such a mask is used downstream from the AO system,
this low throughput does not change the limiting magnitude of the
instrument, solely imposed by the upstream AO performance.

Further work is obviously required to verify the benefits of employing
one such annular aperture mask on an ELT, and will be the object of
publications to come. Nevertheless, this preliminary study means to
show that AO-equipped ELTs can be thought of as interferometers with
some very interesting imaging capabilities, opening access to a realm
of angular separation $\sim5$ mas in the near IR, that projected on a
young association like Taurus (145 pc), allows to probe down to the
inner 1 AU region of potential planet host systems.

\section{Conclusion}

Thinking of the image formation process from an interferometric point
of view offers an alternative to the classical object-image
convolution problem, with some serious advantages.
Using a series of simple examples, the paper shows how the notion of
closure-phase can naturally be generalized when one assumes that
wavefront errors are small ($\lesssim 1$ radian). The convolution
equation can be replaced by a linear model for the phase measured in
the Fourier plane. Kernel-phase, a generalization of closure-phase,
are associated to singular modes of the phase transfer matrix that
propagates pupil phase errors into phase in the Fourier-plane.

Initially envisionned for the processing of narrow-band very
high-Strehl data acquired from space, kernel-phase slowly appears to
be a quite versatile tool, actually compatible with medium and wide
band filters, and quite able to handle reasonably well corrected AO
images.

Thinking ahead about the possibilities offered by the large aperture
of ELTs, one sees a fair bit of potential for very high angular
resolution imaging, using this interferometric framework. Kernel-phase
is a powerful tool for high angular resolution imaging that explores a
still exclusive region of the parameter space.


\begin{thebibliography}{99}
\bibitem{1971JOSA..61..272}
  {Golay}, M. 1971, JOSA, \textbf{61}, 272

\bibitem{2001PASP..113..105H}
  {Hayward} {et~al.} 2001, \pasp, \textbf{113}, 105

\bibitem{2011ApJ...726..104H}
  {Hinkley} {et~al.} 2011, \apj, \textbf{726}, 104

\bibitem{2008ApJ...679..762K}
  {Kraus} {et~al.} 2008, \apj, \textbf{679}, 762

\bibitem{2011ApJ...731....8K}
  {Kraus} {et~al.} 2011, \apj, \textbf{731}, 8

\bibitem{2012ApJ...745....5K}
  {Kraus} \& {Ireland} 2012, \apj, \textbf{745}, 5

\bibitem{2008ApJ...678..463I}
  {Ireland} {et~al.} 2008, \apj, \textbf{678}, 463

\bibitem{2013MNRAS.433.1718I}
  {Ireland}, M.~J. 2013, \mnras, \textbf{433}, 1718

\bibitem{1958MNRAS.118..276J}
  {Jennison}, R.~C. 1958, \mnras, \textbf{118}, 276

\bibitem{2006ApJ...641..556M}
  {Marois} {et~al.} 2006, \apj, \textbf{641}, 556

\bibitem{2010ApJ...724..464M}
  {Martinache}, F. 2010, \apj, \textbf{724}, 464

\bibitem{2012SPIE.8445E..04M}
  {Martinache}, F. 2012, SPIE, \textbf{8445}, 04

\bibitem{2012PASP..124.1288M}
  {Martinache} {et~al.} 2012, \pasp, \textbf{124}, 1288

\bibitem{2000plbs.conf..203M}
  {Monnier} J.~D., 2000, \textbf{Principles of Long Baseline Stellar
    Interferometry}, 203

\bibitem{2013ApJ...767..110P}
  {Pope} {et~al.} 2013, \apj, \textbf{767}, 110

\bibitem{2006SPIE.6272E.103T}
  {Tuthill} {et~al.} 2006, SPIE, \textbf{6272}, 103

\bibitem{2008ApJ...675..698T}
  {Tuthill} {et~al.} 2008, \apj, \textbf{675}, 698

\end{thebibliography}

\end{document}